\documentclass[3p,times,twocolumn]{elsarticle}

\usepackage{ecrc}

\usepackage[utf8x]{inputenc}
\usepackage{amsmath}
\usepackage[caption=false]{subfig}

\volume{00}

\firstpage{1}

\journalname{Nuclear Physics B Proceedings Supplement}

\runauth{}

\jid{nppp}

\jnltitlelogo{Nuclear Physics B Proceedings Supplement}

\usepackage{amssymb}

\usepackage[figuresright]{rotating}

\usepackage{hyperref}

\newcommand{\der}{\mathrm{d}}

\newcommand{\tr}{\, \mathrm{Tr} \, }

\newcommand{\nc}{{N_\mathrm{c}}}
\newcommand{\nf}{{n_\mathrm{f}}}

\newcommand{\qs}{Q_\mathrm{s}}

\newcommand{\qso}{Q_\mathrm{s,0}}

\newcommand{\lqcd}{\Lambda_{\mathrm{QCD}}}
\newcommand{\as}{\alpha_{\mathrm{s}}}

\begin{document}

\begin{frontmatter}



\dochead{}

\title{Solving the Balitsky-Kovchegov equation at next to leading order accuracy}


\author[jyu,hip]{T. Lappi}
\author[jyu]{H. Mäntysaari}
\address[jyu]{Department of Physics, University of Jyväskylä, P.O. Box 35, 40014 University of Jyväskylä, Finland}
\address[hip]{Helsinki Institute of Physics, P.O. Box 64, 00014 University of Helsinki, Finland}

\begin{abstract}
We solve the Balitsky-Kovchegov small-$x$ evolution equation in coordinate space. We find that the solution to the equation is unstable when using an initial condition relevant for phenomenological applications at leading order. The problematic behavior is shown to be due to a large double logarithmic contribution. The same problem is found when the evolution of the ``conformal dipole'' is solved, even though the double logarithmic term is then absent from the evolution equation.
\end{abstract}

\begin{keyword}
BK, DIS, CGC


\end{keyword}

\end{frontmatter}


\section{Introduction}
\label{}
The Color Glass Condensate \cite{Gelis:2010nm} effective theory of QCD at high energy has been shown to be in good agreement with large amount of experimental data on, for example, deep inelastic scattering, single and double inclusive particle production and exclusive vector meson production, see e.g.~\cite*{Lappi:2012nh,Lappi:2013zma,Lappi:2013am}. There are two main ingredients in these calculations: first, one needs the small-$x$ evolution equation such as the Balitsky-Kovchegov (BK) equation~\cite{Balitsky:1995ub,Kovchegov:1999yj} which describes the evolution of the dipole-target scattering amplitude as a function of energy, or equivalently, Bjorken-$x$. The dipole amplitude at initial Bjorken-$x$ is the second necessary input, and it can not be obtained from perturbative calculations but it must be fit to experimental data. The fact that one can indeed obtain a good description of the precise combined HERA single inclusive data~\cite{Aaron:2009aa} has been one of the tightest tests for the CGC~\cite{Lappi:2013zma,Albacete:2010sy}.

An important next step for the description of the saturation phenomena from the CGC framework is to develop the CGC theory to the next to leading order accuracy. First steps in this direction have been taken by deriving e.g. the photon impact factor~\cite{Beuf:2011xd} and single inclusive cross section~\cite{Chirilli:2012jd} at this order. The NLO BK equation is also known~\cite{Balitsky:2008zza}, but no solution to it existed before our work~\cite{Lappi:2015fma}.

\section{The BK equation at NLO}
The BK evolution equation for the dipole operator $S$, which is a correlator of Wilson lines $U$ such that $S(x-y)=1/\nc \langle \tr  U^\dagger(x)U(y)\rangle$. At NLO accuracy the equation reads
\begin{multline}
\label{eq:nlobk}
	\partial_y S(r) = \frac{\as \nc}{2\pi^2} K_1 \otimes [S(X)S(Y)-S(r)] \\
		+ \frac{\as^2 \nc^2}{8\pi^4} K_2 \otimes [S(X)S(z-z')S(Y')-S(X)S(Y)] \\
		+ \frac{\as^2 \nf \nc}{8\pi^4} K_f \otimes S(Y)[S(X')-S(X)].
\end{multline} 
The quark and the antiquark of the parent dipole are at transverse positions $x$ and $y$, and the daughter dipole sizes are $X=|x-z|$, $Y=|y-z|$, $X'=|x-z'|$ and $Y'=|y-z'|$, and $r$ is the size of the parent dipole. The convolutions $\otimes$ are taken by integrating over the daughter dipole sizes ($z$ in $K_1$ and both $z$ and $z'$ in $K_2$ and $K_f$).

The kernel $K_1$ includes the leading order BK kernel, the running coupling part and an $\as^2$ correction, as
\begin{multline}
K_1 = \frac{r^2}{X^2Y^2} \left[ 1+\frac{\as\nc }{4\pi} \left(  \frac{\beta}{\nc} \ln r^2\mu^2 - \frac{\beta}{\nc} \frac{X^2-Y^2}{r^2} \ln \frac{X^2}{Y^2} \right. \right. \\
 \left. \left. + \frac{67}{9} - \frac{\pi^2}{3} - \frac{10}{9} \frac{\nf}{\nc} - \ln \frac{X^2}{r^2} \ln \frac{Y^2}{r^2} \right) \right].
\end{multline}
We implement the running coupling by replacing the terms proportional to the $\beta$ function coefficient by the Balitsky running coupling prescription from Ref.~\cite{Balitsky:2006wa}. The kernel $K_1$ then reads
\begin{multline} 
\frac{\as \nc}{2\pi^2} K_1 = \frac{\as(r) \nc}{2\pi^2} \\
\times  \left[\frac{r^2}{X^2Y^2} + \frac{1}{X^2} \left(\frac{\as(X)}{\as(Y)}-1\right) + \frac{1}{Y^2} \left(\frac{\as(Y)}{\as(X)}-1\right) \right] \\
		+ \frac{\as(r)^2 \nc^2}{8\pi^3} \frac{r^2}{X^2Y^2} \left[ \frac{67}{9} - \frac{\pi^2}{3} - \frac{10}{9} \frac{\nf}{\nc} - 2\ln \frac{X^2}{r^2} \ln \frac{Y^2}{r^2} \right].
\end{multline}
We will later refer to the part proportional to $\ln X^2/r^2 \ln Y^2/r^2$ as the double logarithmic term.

The kernels $K_2$ and $K_f$ are combinations of rational expressions of transverse separations and a logarithm $\ln X^2Y'^2/(X'^2Y^2)$.   Note that this logarithm vanishes in the small parent dipole limit where $x\to y$, in contrast to the double logarithm. The coupling constant $\as$ is evaluated at the scale set by the parent dipole, as it is the only external scale. For explicit expressions, we refer the reader to Refs.~\cite{Balitsky:2008zza,Lappi:2015fma}.

As an initial condition for the NLO BK equation we use a modified McLerran-Venugopalan (MV) model
\begin{equation}
	N(r) = 1 - S(r) = 1- \exp \left[ -\frac{(r^2\qso^2)^\gamma}{4} \ln \left(\frac{1}{r\lqcd}+e \right) \right].
\end{equation}
Here the MV model is modified by introducing an anomalous dimension $\gamma$ which controls the power-like tail of the dipole amplitude at small dipole sizes. The leading order fits to the HERA data prefer~\cite{Albacete:2010sy} values of $\gamma\sim 1.1$, which then reduces during the evolution to $\gamma\sim 0.8$. The constant $\qso$ parametrizes the saturation scale at initial Bjorken-$x$. In this work, we do not seek for parameter values that are compatible with the experimental data. In practice, $\qso$ controls the relative importance of the NLO terms as it scales the value of $\as$.

\section{Solution to the NLO BK}

\begin{figure}[tb]
\begin{center}
\includegraphics[width=0.49\textwidth]{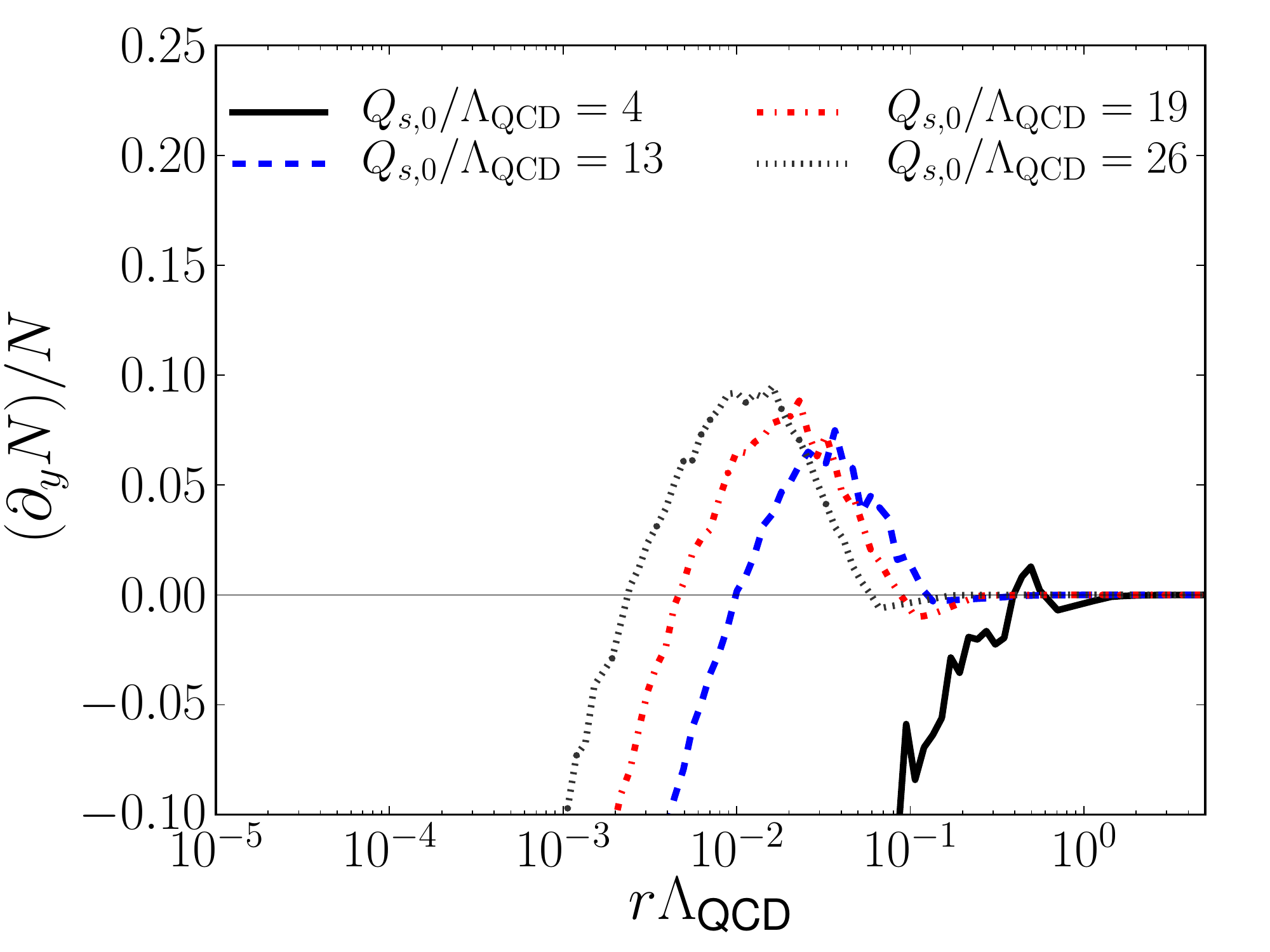}
\end{center}
\caption{
Evolution speed of the dipole amplitude at initial condition (MV model with $\gamma=1$) with different values for the initial saturation scale.
}\label{fig:dndy_qcd_y0}
\end{figure}

In Fig.~\ref{fig:dndy_qcd_y0} we show the evolution speed $\partial_y N(r)/N(r)$ as a function of the dipole size for the MV model ($\gamma=1$) initial condition. At small initial saturation scales $\qso/\lqcd$, when the strong coupling constant and the NLO corrections are largest, the evolution speed is negative at all dipole sizes. With smaller values of $\as$ (larger saturation scale) the evolution speed turns negative at small dipole sizes when $r \ll 1/Q_s$.

The negative evolution speed is unphysical, as it corresponds to having an unintegrated gluon distribution that decreases when it is probed at smaller and smaller $x$. However, having $\partial_y N/N \sim \ln r$ in the small $r$ limit is a signal of mathematical instability, as in that case there is a small (but finite) $r$ below which the dipole amplitude becomes negative in one step $\der y$ of the rapidity evolution. On the other hand, the definition of the dipole amplitude $N(x-y)=1-1/\nc \langle \tr U^\dagger(x)U(y) \rangle$ requires that $N(r)\to 0$ in the limit $r\to 0$. Also, if the dipole amplitude does not satisfy this requirement the $z$ integral in the leading order equation does not converge. In our numerical analysis we impose by hand a constraint $N(r)\ge 0$.

\begin{figure}[tb]
\begin{center}
\includegraphics[width=0.49\textwidth]{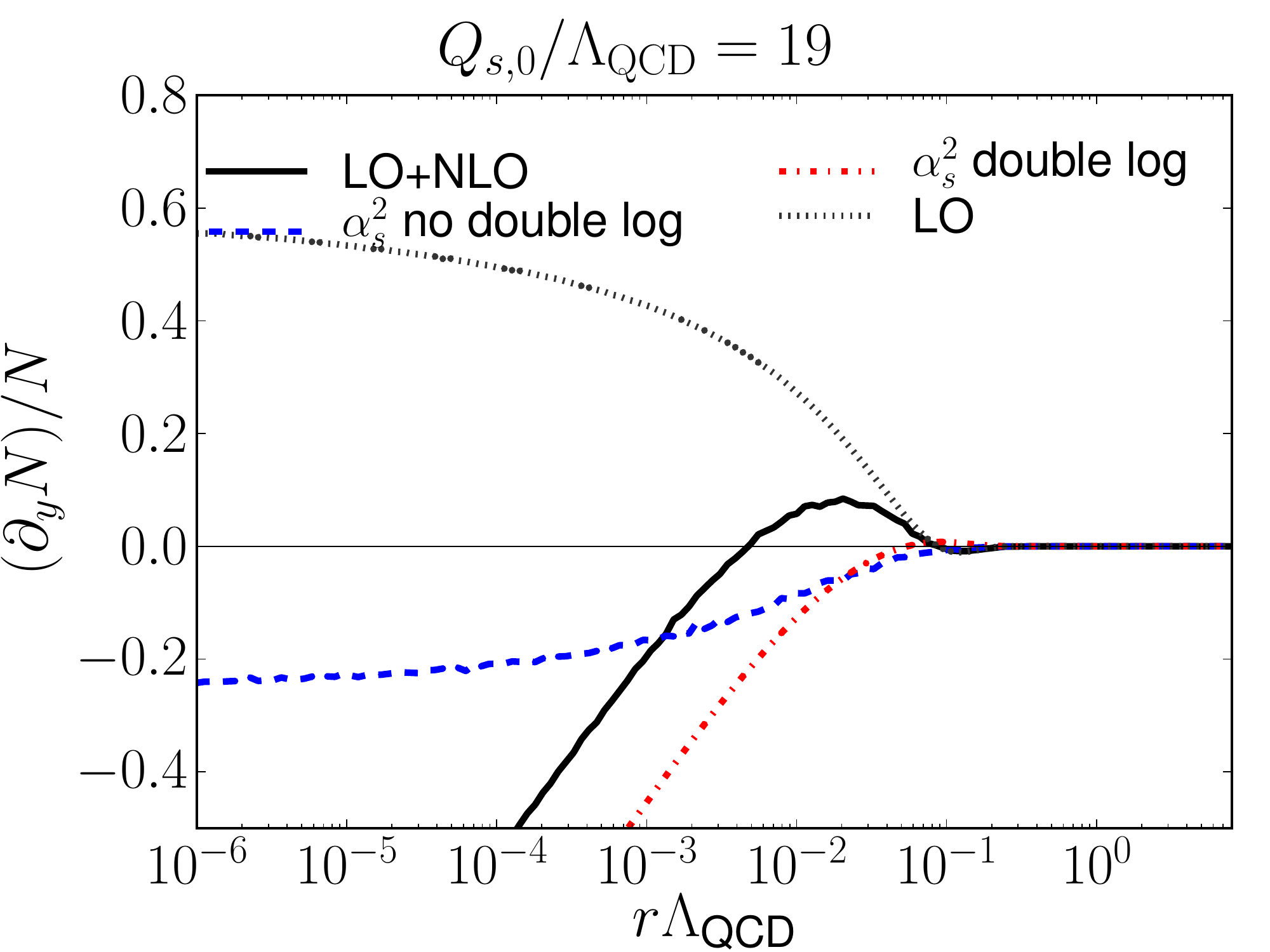}
\end{center}
\caption{
Contributions to the evolution speed of the dipole amplitude originating from different terms of the NLO BK equation at the initial condition (MV model with $\gamma=1$).
}\label{fig:dndy_qcd_terms}
\end{figure}

Let us then trace back the origin of the negative evolution speed. In Fig.~\ref{fig:dndy_qcd_terms} we show contributions to $\partial_y N/N$ originating from the  different terms of the NLO BK equation. We observe that the double logarithmic term, which is part of the kernel $K_1$, is the one that drives the evolution speed negative. The other NLO corrections also decrease the evolution speed but do not cause the problematic $\partial_y N/N \sim \ln r$ behavior.

\begin{figure}[tb]
\begin{center}
\includegraphics[width=0.49\textwidth]{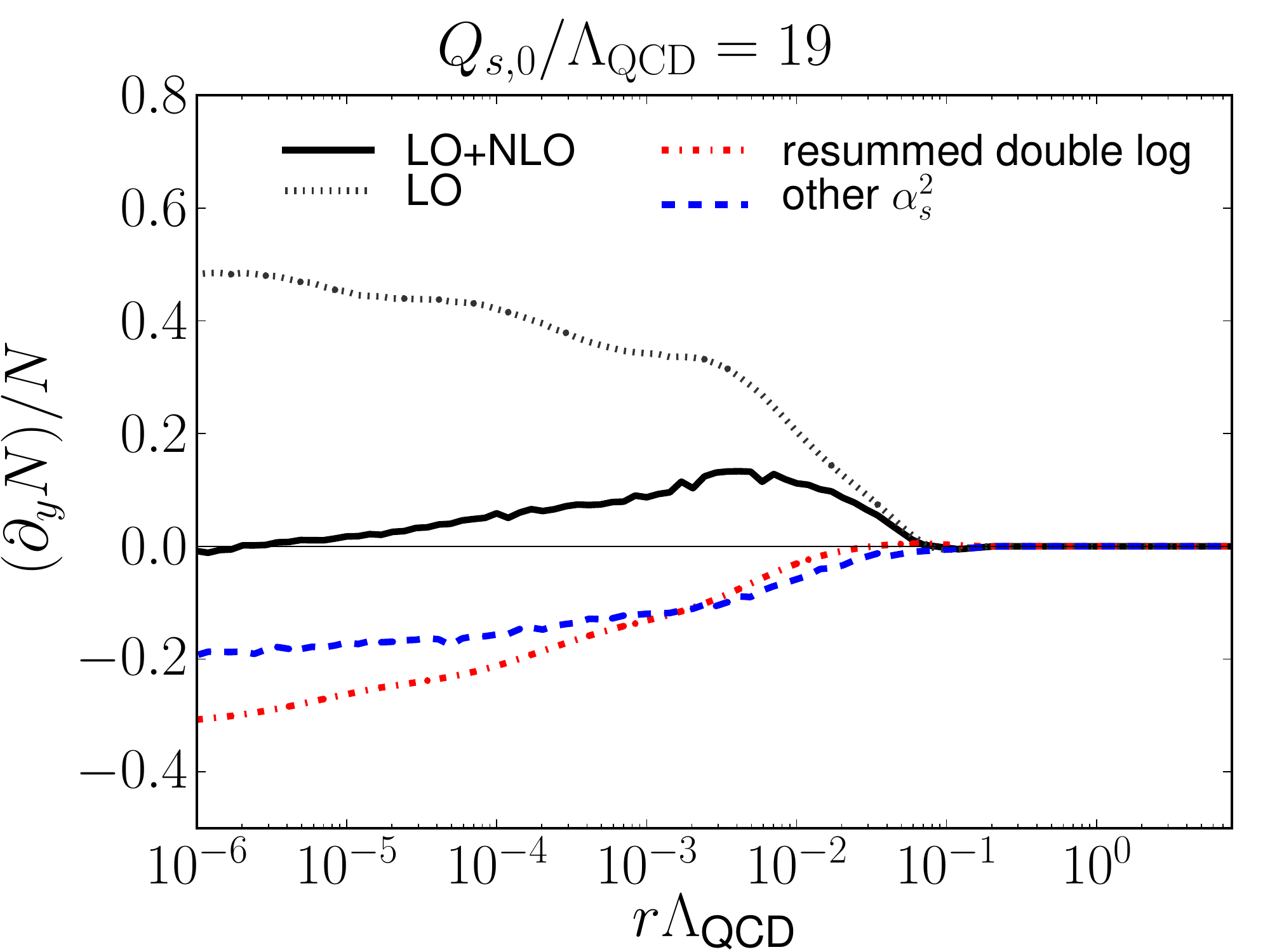}
\end{center}
\caption{
Evolution speed of the dipole amplitude obtained by resumming the double logarithmic terms to all orders. Result is shown at initial condition (MV model with $\gamma=1$), and the contributions from the resummation and from the other NLO terms are shown separately.
}\label{fig:dndy_qcd_resum}
\end{figure}

It has been argued in Ref.~\cite{Iancu:2015vea} that the double logarithmic contributions should be resummed to all orders. The resummation effectively removes the double logarithmic term from the kernel $K_1$ and multiplies the leading order BK kernel $r^2/X^2Y^2$ by an oscillatory factor, which expanded to order $\as^2$ gives the double logarithmic term to the kernel $K_1$. The initial condition is also modified by the resummation procedure.  We implement this resummation in our analysis and show in Fig.~\ref{fig:dndy_qcd_resum} the evolution speed obtained by solving the NLO BK equation with  resummation. We find that when the double logarithmic corrections are resummed to all orders the evolution speed turns negative at significantly smaller dipoles, but the negativity problem is not completely solved. However, it was later shown in Ref.~\cite{Iancu:2015joa} that there is also a single log term that should be resummed.

\begin{figure*}[ptb]
	\subfloat[$y=1$]{
		\includegraphics[width=0.33\textwidth]{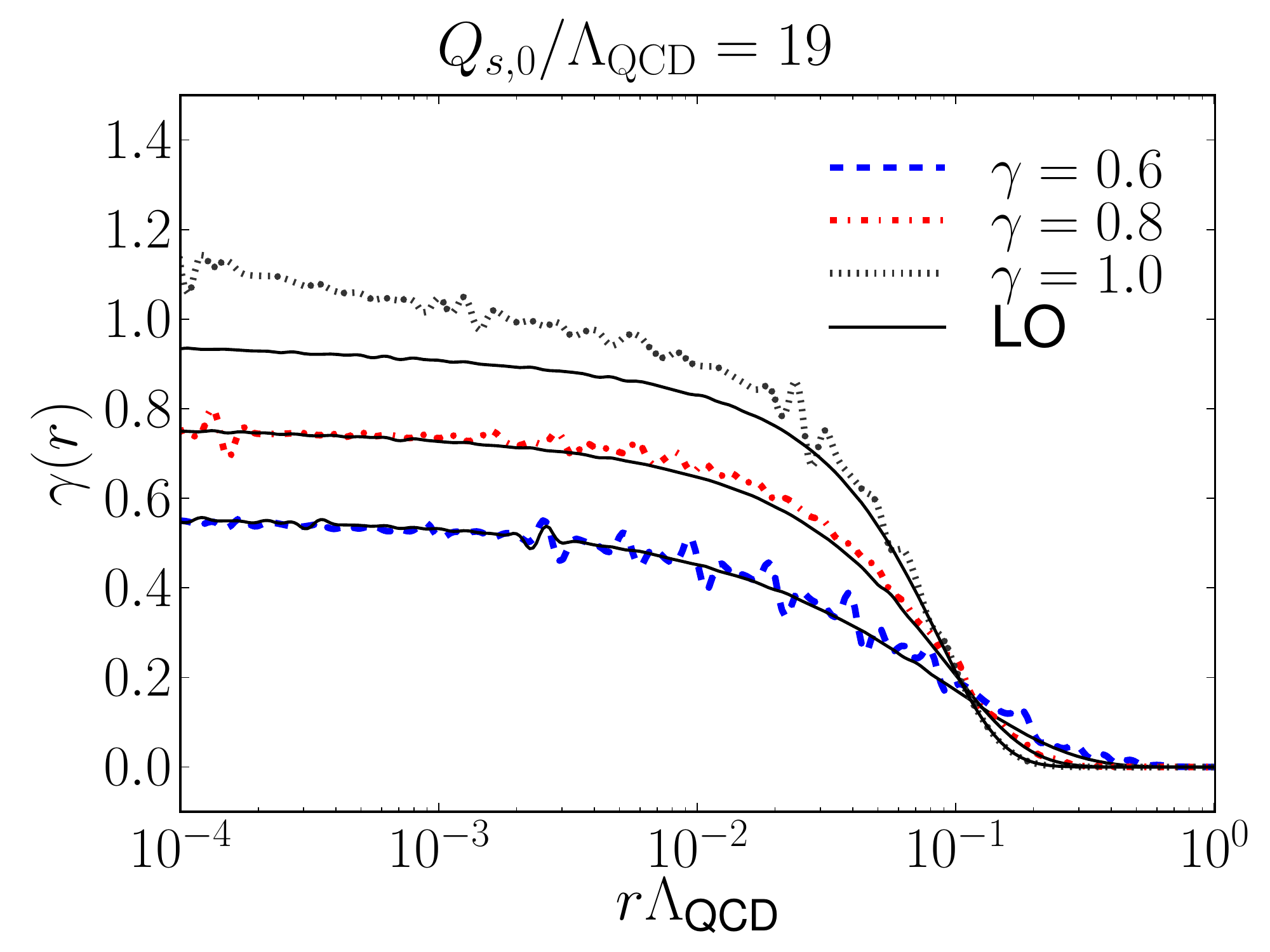}
		}
	\subfloat[$y=5$]{
		\includegraphics[width=0.33\textwidth]{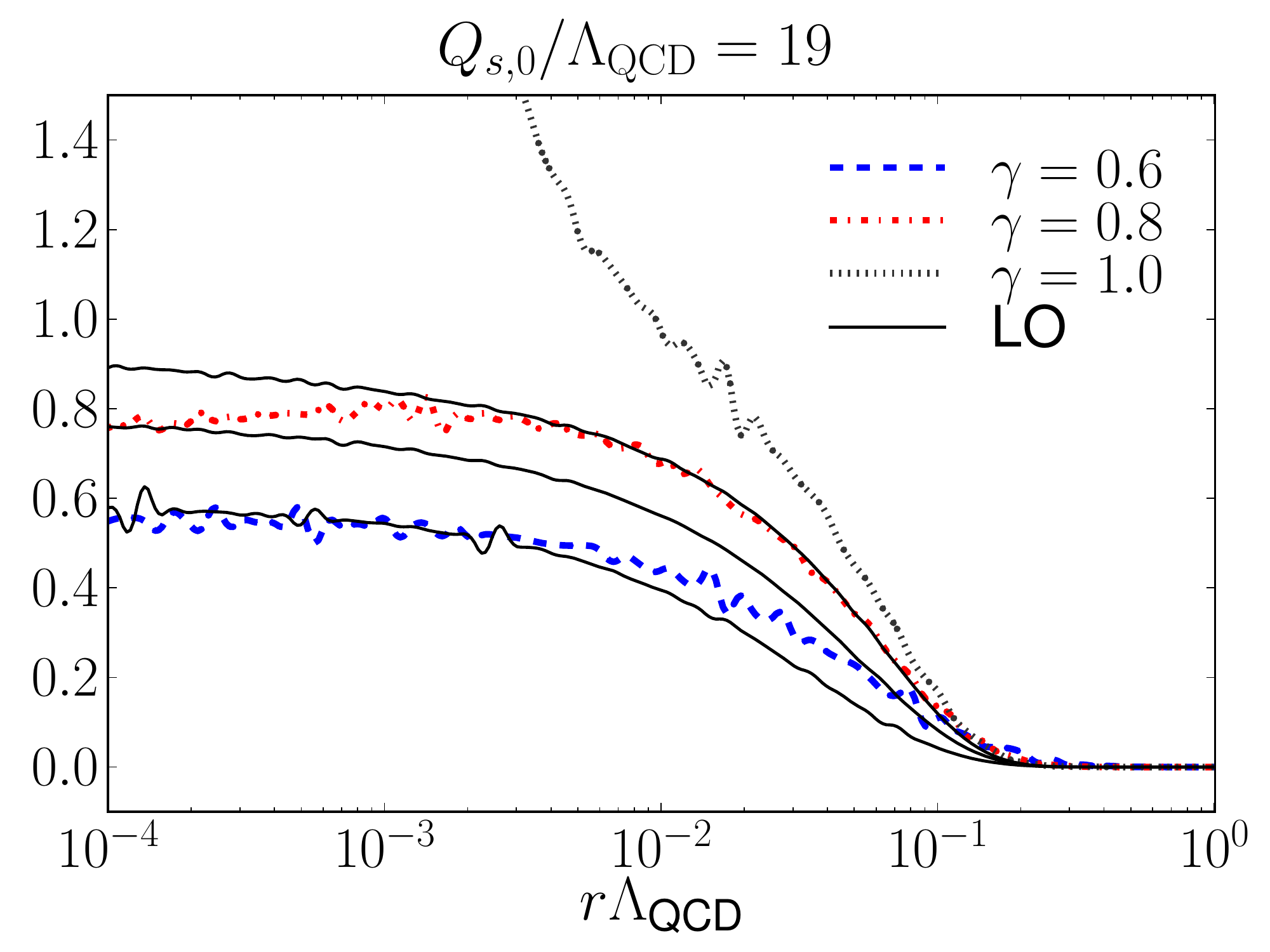}
		}
	\subfloat[$y=30$]{
		\includegraphics[width=0.33\textwidth]{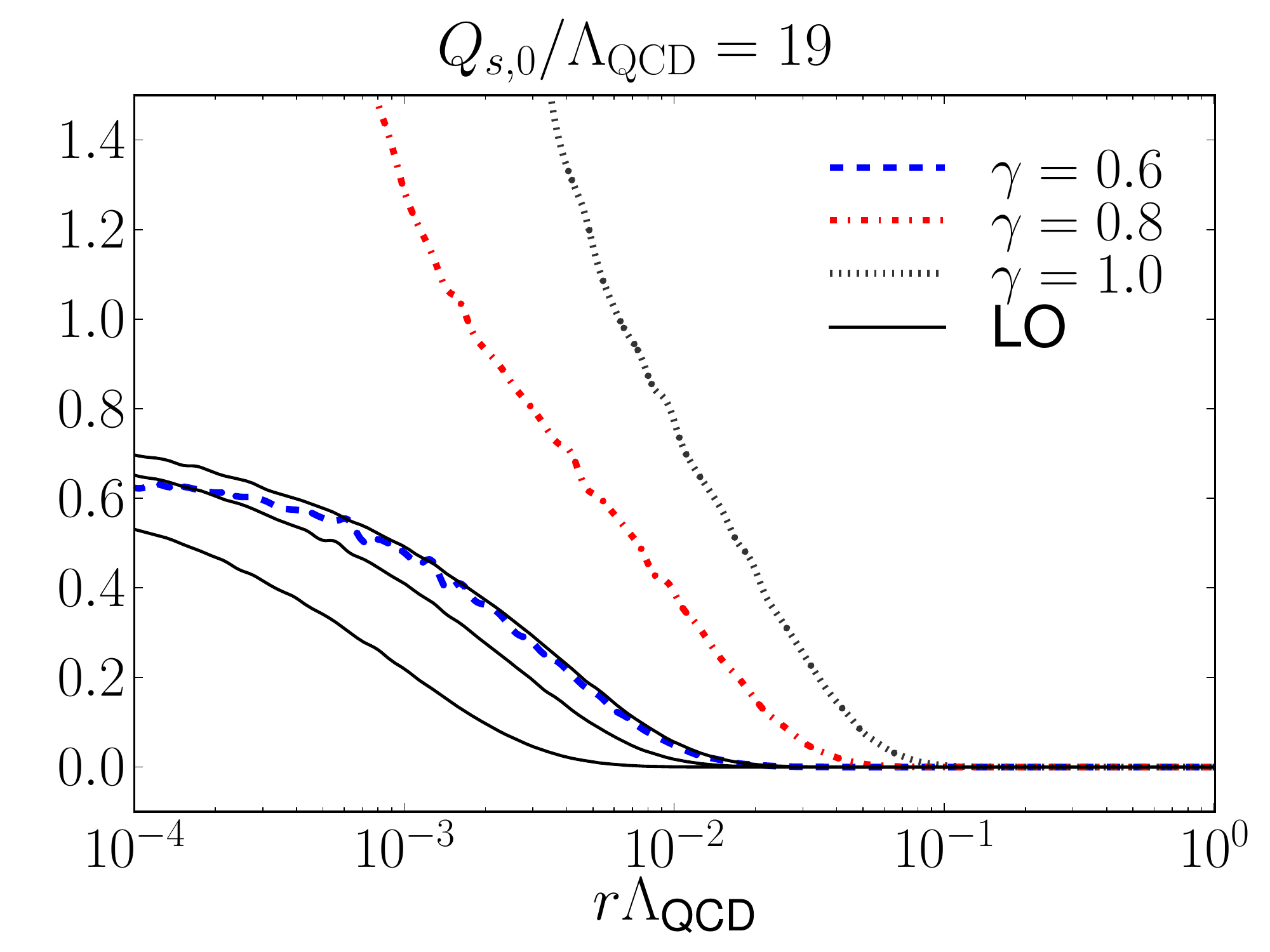}
		}			
	\caption{Anomalous dimension $\gamma(r)$ at rapidities $y=1,5,30$ with different values for parameter $\gamma$ in the initial condition.}
	\label{fig:gammaqcd}
\end{figure*}

Let us then study the dependence on the anomalous dimension $\gamma$. We define the dipole size dependent anomalous dimension as $\gamma(r) = \der \ln N(r)/\der \ln r^2$ and show it in Fig.~\ref{fig:gammaqcd} at different rapidities using different values for the parameter $\gamma$ in the initial condition. With $\gamma=1$ in the initial condition, the anomalous dimension increases rapidly in the evolution, which is a signal of the dipole amplitude approaching a step function form $N(r)\sim \theta(r-1/\qs)$ (recall that we force $N(r)\ge 0$). This unstable behavior is not seen with $\gamma=0.6$ in the initial condition within the rapidity range studied here. Indeed, with $\gamma=0.6$ the evolution speed actually remains positive at small dipoles ($r\lqcd \gtrsim 10^{-6}$ included in the numerical calculation) within the studied rapidity interval.

\section{Conformal dipole}

The Wilson lines are by definition conformally invariant, and thus one expects the evolution equation for the dipole operator $S$ to also respect this symmetry if the running of the coupling is neglected. However, in the NLO BK equation the double logarithmic term, which is responsible for the unstable behavior, is not conformal. This non-conformal term also appears in the evolution equation derived in $N=4$ Super Yang-Mills theory, as shown in Ref.~\cite{Balitsky:2009xg}. It is therefore interpreted as an artefact of a cutoff that breaks the conformal invariance.

In Ref.~\cite{Balitsky:2009xg} the authors present a way to restore the conformal invariance by writing the evolution equation for a \emph{conformal dipole} defined as
\begin{multline}
	S(r)^\text{conf} = S(r) - \frac{\as \nc}{4\pi^2} \int \der^2 z \frac{r^2}{X^2Y^2} \ln \frac{ar^2}{X^2Y^2} \\
	\times [ S(X)S(Y)-S(r)].
\end{multline}
Here $a$ is a dimensionful constant that drops out of the evolution equation. Writing the evolution equation for the conformal dipole removes the double logarithmic term from $K_1$ and introduces a new term proportional to $\ln r^2$ in $K_2$. We have solved the NLO BK equation for the conformal dipole, and as a result find that qualitatively the evolution does not differ from the evolution of the ``non-conformal'' dipole. As shown in Ref.~\cite{Lappi:2015fma}, the new $\ln r^2$ logarithm now drives the evolution speed negative at small dipoles.

\section{Discussion}
We have obtained the first numerical solution to the Balitsky-Kovchegov equation at next to leading order, and the NLO corrections are shown in decrease the evolution speed. This is expected, as the leading order fits to the deep inelastic scattering data seem to favor relatively small evolution speeds. The double logarithmic term in the NLO BK equation, however, causes an unstable behavior at small dipoles when an initial condition relevant to phenomenological applications (at least at leading order) is used. The resummation procedures shown in Refs.~\cite{Iancu:2015vea,Iancu:2015joa} have potential to improve the situation and bring the NLO BK equation closer to the point where it can be used in phenomenological applications. In addition to the resummation schemes, also the role of the kinematical constraint to the BK equation~\cite{Motyka:2009gi,Beuf:2014uia} should be better understood.

\section*{Acknowledgements}
This work has been supported by the Academy of Finland, projects 267321
and 273464, the Graduate School of Particle and Nuclear Physics (H.M.) and by computing resources from CSC -- IT Center for Science in Espoo, Finland.



\bibliographystyle{elsarticle-num}
\bibliography{../../refs-elsevier}

\begin{thebibliography}{10}
\expandafter\ifx\csname url\endcsname\relax
  \def\url#1{\texttt{#1}}\fi
\expandafter\ifx\csname urlprefix\endcsname\relax\def\urlprefix{URL }\fi
\expandafter\ifx\csname href\endcsname\relax
  \def\href#1#2{#2} \def\path#1{#1}\fi

\bibitem{Gelis:2010nm}
F.~Gelis, E.~Iancu, J.~Jalilian-Marian, R.~Venugopalan, {The Color Glass
  Condensate}, Ann. Rev. Nucl. Part. Sci. 60 (2010) 463--489.
\newblock \href {http://arxiv.org/abs/1002.0333} {\path{arXiv:1002.0333}},
  \href {http://dx.doi.org/10.1146/annurev.nucl.010909.083629}
  {\path{doi:10.1146/annurev.nucl.010909.083629}}.

\bibitem{Lappi:2012nh}
T.~Lappi, H.~M{\"a}ntysaari, {Forward dihadron correlations in deuteron-gold
  collisions with the Gaussian approximation of JIMWLK}, Nucl.Phys. A908 (2013)
  51--72.
\newblock \href {http://arxiv.org/abs/1209.2853} {\path{arXiv:1209.2853}},
  \href {http://dx.doi.org/10.1016/j.nuclphysa.2013.03.017}
  {\path{doi:10.1016/j.nuclphysa.2013.03.017}}.

\bibitem{Lappi:2013zma}
T.~Lappi, H.~Mäntysaari, {Single inclusive particle production at high energy
  from HERA data to proton-nucleus collisions}, Phys. Rev. D88 (2013) 114020.
\newblock \href {http://arxiv.org/abs/1309.6963} {\path{arXiv:1309.6963}},
  \href {http://dx.doi.org/10.1103/PhysRevD.88.114020}
  {\path{doi:10.1103/PhysRevD.88.114020}}.

\bibitem{Lappi:2013am}
T.~Lappi, H.~M{\"a}ntysaari, {$J/\Psi$ production in ultraperipheral Pb+Pb and
  p+Pb collisions at LHC energies}, Phys. Rev. C87 (2013) 032201.
\newblock \href {http://arxiv.org/abs/1301.4095} {\path{arXiv:1301.4095}},
  \href {http://dx.doi.org/10.1103/PhysRevC.87.032201}
  {\path{doi:10.1103/PhysRevC.87.032201}}.

\bibitem{Balitsky:1995ub}
I.~Balitsky, {Operator expansion for high-energy scattering}, Nucl. Phys. B463
  (1996) 99--160.
\newblock \href {http://arxiv.org/abs/hep-ph/9509348}
  {\path{arXiv:hep-ph/9509348}}, \href
  {http://dx.doi.org/10.1016/0550-3213(95)00638-9}
  {\path{doi:10.1016/0550-3213(95)00638-9}}.

\bibitem{Kovchegov:1999yj}
Y.~V. Kovchegov, {Small-$x$ $F_2$ structure function of a nucleus including
  multiple pomeron exchanges}, Phys. Rev. D60 (1999) 034008.
\newblock \href {http://arxiv.org/abs/hep-ph/9901281}
  {\path{arXiv:hep-ph/9901281}}, \href
  {http://dx.doi.org/10.1103/PhysRevD.60.034008}
  {\path{doi:10.1103/PhysRevD.60.034008}}.

\bibitem{Aaron:2009aa}
F.~Aaron, et~al., {Combined Measurement and QCD Analysis of the Inclusive
  $e^\pm p$ Scattering Cross Sections at HERA}, JHEP 1001 (2010) 109.
\newblock \href {http://arxiv.org/abs/0911.0884} {\path{arXiv:0911.0884}},
  \href {http://dx.doi.org/10.1007/JHEP01(2010)109}
  {\path{doi:10.1007/JHEP01(2010)109}}.

\bibitem{Albacete:2010sy}
J.~L. Albacete, N.~Armesto, J.~G. Milhano, P.~Quiroga-Arias, C.~A. Salgado,
  {AAMQS: A non-linear QCD analysis of new HERA data at small-x including heavy
  quarks}, Eur. Phys. J. C71 (2011) 1705.
\newblock \href {http://arxiv.org/abs/1012.4408} {\path{arXiv:1012.4408}},
  \href {http://dx.doi.org/10.1140/epjc/s10052-011-1705-3}
  {\path{doi:10.1140/epjc/s10052-011-1705-3}}.

\bibitem{Beuf:2011xd}
G.~Beuf, {NLO corrections for the dipole factorization of DIS structure
  functions at low x}, Phys. Rev. D85 (2012) 034039.
\newblock \href {http://arxiv.org/abs/1112.4501} {\path{arXiv:1112.4501}},
  \href {http://dx.doi.org/10.1103/PhysRevD.85.034039}
  {\path{doi:10.1103/PhysRevD.85.034039}}.

\bibitem{Chirilli:2012jd}
G.~A. Chirilli, B.-W. Xiao, F.~Yuan, {Inclusive Hadron Productions in pA
  Collisions}, Phys. Rev. D86 (2012) 054005.
\newblock \href {http://arxiv.org/abs/1203.6139} {\path{arXiv:1203.6139}},
  \href {http://dx.doi.org/10.1103/PhysRevD.86.054005}
  {\path{doi:10.1103/PhysRevD.86.054005}}.

\bibitem{Balitsky:2008zza}
I.~Balitsky, G.~A. Chirilli, {Next-to-leading order evolution of color
  dipoles}, Phys. Rev. D77 (2008) 014019.
\newblock \href {http://arxiv.org/abs/0710.4330} {\path{arXiv:0710.4330}},
  \href {http://dx.doi.org/10.1103/PhysRevD.77.014019}
  {\path{doi:10.1103/PhysRevD.77.014019}}.

\bibitem{Lappi:2015fma}
T.~Lappi, H.~Mäntysaari, {Direct numerical solution of the coordinate space
  Balitsky-Kovchegov equation at next to leading order}, Phys. Rev. D91 (2015)
  074016.
\newblock \href {http://arxiv.org/abs/1502.02400} {\path{arXiv:1502.02400}},
  \href {http://dx.doi.org/10.1103/PhysRevD.91.074016}
  {\path{doi:10.1103/PhysRevD.91.074016}}.

\bibitem{Balitsky:2006wa}
I.~Balitsky, {Quark contribution to the small-x evolution of color dipole},
  Phys. Rev. D75 (2007) 014001.
\newblock \href {http://arxiv.org/abs/hep-ph/0609105}
  {\path{arXiv:hep-ph/0609105}}, \href
  {http://dx.doi.org/10.1103/PhysRevD.75.014001}
  {\path{doi:10.1103/PhysRevD.75.014001}}.

\bibitem{Iancu:2015vea}
E.~Iancu, J.~Madrigal, A.~Mueller, G.~Soyez, D.~Triantafyllopoulos, {Resumming
  double logarithms in the QCD evolution of color dipoles}, Phys. Lett. B744
  (2015) 293--302.
\newblock \href {http://arxiv.org/abs/1502.05642} {\path{arXiv:1502.05642}},
  \href {http://dx.doi.org/10.1016/j.physletb.2015.03.068}
  {\path{doi:10.1016/j.physletb.2015.03.068}}.

\bibitem{Iancu:2015joa}
E.~Iancu, J.~D. Madrigal, A.~H. Mueller, G.~Soyez, D.~N. Triantafyllopoulos,
  {Collinearly-improved BK evolution meets the HERA data}\href
  {http://arxiv.org/abs/1507.03651} {\path{arXiv:1507.03651}}.

\bibitem{Balitsky:2009xg}
I.~Balitsky, G.~A. Chirilli, {NLO evolution of color dipoles in $N=4$ SYM},
  Nucl. Phys. B822 (2009) 45--87.
\newblock \href {http://arxiv.org/abs/0903.5326} {\path{arXiv:0903.5326}},
  \href {http://dx.doi.org/10.1016/j.nuclphysb.2009.07.003}
  {\path{doi:10.1016/j.nuclphysb.2009.07.003}}.

\bibitem{Motyka:2009gi}
L.~Motyka, A.~M. Stasto, {Exact kinematics in the small $x$ evolution of the
  color dipole and gluon cascade}, Phys. Rev. D79 (2009) 085016.
\newblock \href {http://arxiv.org/abs/0901.4949} {\path{arXiv:0901.4949}},
  \href {http://dx.doi.org/10.1103/PhysRevD.79.085016}
  {\path{doi:10.1103/PhysRevD.79.085016}}.

\bibitem{Beuf:2014uia}
G.~Beuf, {Improving the kinematics for low-x QCD evolution equations in
  coordinate space}, Phys. Rev. D89 (2014) 074039.
\newblock \href {http://arxiv.org/abs/1401.0313} {\path{arXiv:1401.0313}},
  \href {http://dx.doi.org/10.1103/PhysRevD.89.074039}
  {\path{doi:10.1103/PhysRevD.89.074039}}.

\end{thebibliography}







\end{document}